\documentclass[showpacs,superscriptaddress,aps,epsfig,rotate,rotating,pre]{revtex4}
\usepackage[final]{graphicx}

\begin{document}

\title{Study of the Nonequilibrium Critical Quenching and the Annealing Dynamics
for the Long-Range Ising Model in 1-dimension}

\author{D. E. Rodriguez}
\affiliation{Instituto de F{\'i}sica de L{\'i}quidos y Sistemas Biol\'{o}gicos (IFLYSIB),
calle 59 nro 789 (1900), Universidad Nacional de La Plata, CCT-La Plata CONICET, La Plata, Argentina.}
\author{M. A. Bab}
\affiliation{Instituto de Investigaciones Fisicoqu\'{i}micas 
Te\'{o}ricas y Aplicadas (INIFTA), Facultad de Ciencias Exactas, Universidad 
Nacional de La Plata, CCT-La Plata CONICET; Suc. 4, CC16 (1900) La Plata, Argentina.}
\author{E. V. Albano}
\affiliation{Instituto de F{\'i}sica de L{\'i}quidos y Sistemas Biol\'{o}gicos, Facultad de Ciencias
Exactas, Universidad Nacional de La Plata, CCT-La Plata CONICET, La Plata, Argentina.}

\begin{abstract}
Extensive Monte Carlo simulations are employed in order to study the dynamic
critical behaviour of the one-dimensional Ising magnet, with algebraically
decaying long-range interactions of the form $\frac{1}{r^{d+\sigma}}$, 
with $\sigma=0.75$. The critical temperature, as well as the critical 
exponents, are evaluated from the power-law behaviour of suitable 
physical observables when the system is quenched from uncorrelated 
states, corresponding to infinite temperature, to the critical point. 
These results are compared with those obtained from the dynamic evolution 
of the system when it is suddenly annealed at the critical point from 
the ordered state. Also, the critical temperature in the infinite 
interaction limit is obtained by means of a finite-range scaling 
analysis of data measured with different truncated-interaction range. 
All the estimated static critical exponents ($\gamma /\nu $, $\beta /\nu $, and 
$1/\nu $ ) are in good agreement with Renormalization Group (RG)
results and previously reported numerical data obtained under equilibrium
conditions. On the other hand, the dynamic exponent of the initial increase of the magnetization ($\theta$) was 
close to RG predictions. However, the dynamic 
exponent $z$ of the time correlation length is slightly different than the RG results 
likely due to the fact that either it may depend on the specific dynamics used 
or because the two-loop expansion used in the RG analysis may be insufficient.

\end{abstract}

\maketitle

\section{Introduction}

The study of the critical behaviour of systems with long-range (LR)
interactions is still a challenging topic in the field of
statistical physics \cite{chen,lut,ber,fish}. Furthermore, the understanding of
the dynamic evolution of these systems, from far-from-equilibrium initial
states towards a final equilibrium regime, poses an additional difficulty
due to the fast relaxation of relevant physical observables owing to the
presence of LR interactions. For these reasons, the study of relaxation
processes in simple Ising and Potts models with LR interactions plays an
important role for the understanding of the dynamics of second-order phase
transitions. Within this context, the study of the short-time dynamics (STD)
of critical systems has attracted great attention during the last two
decades \cite{Zheng,Oliveira,chen,loca}, for a recent review see e.g. \cite{RPP}. The pioneering theoretical study of
the STD, which was formulated in the context of the dynamic
Renormalization Group \cite{jan}, predicts the existence of a new exponent
related to the initial increase of the order parameter. This prediction has
subsequently been validated by a large body of numerical evidence obtained
in a variety of models \cite{Zheng,RPP,bab,bab2,santos,silva,grandi}. However,
only few studies have been performed in order to generalize these concepts
to systems with LR interactions. In fact, the field-theoretical calculations
of Janssen et al. \cite{jan} have been extended to the case of LR interactions
decaying according to a power law for the case of the continuous $n-$vector
model \cite{chen}, the random Ising model \cite{chen1}, and the kinetic
spherical model \cite{chen2,bau}. On the other hand, theoretical studies of
the relaxation dynamics of discrete models are still lacking, and only few
preliminary numerical results on the STD of the Potts model have recently
been reported \cite{loca}.

In order to contribute to the understanding of the dynamics of phase
transitions in discrete systems, the aim of this paper is to report and
discuss extensive numerical simulations of the Ising model, in one
dimension, with LR interactions decaying with the distance as a power law.
For this purpose, we performed studies of both the STD of initially
disordered states (i.e., quenching experiments) and the relaxation dynamics
of initially ordered states (i.e., annealing experiments). Results obtained
by applying these methods allow us to determine not only the critical
temperature, but also the complete set of static and dynamic critical
exponents (for the methodology used, see e.g. \cite{RPP,gab_bag}). In this way, we
can compare our results with theoretical Renormalization Group (RG) results
\cite{chen,Binder} and with independent numerical determinations of the
static exponents performed under equilibrium conditions \cite{ber}.

The paper is organized as follows: in Section II a brief description of the
model and the simulation method is presented, Section III is devoted to a
brief discussion of the theoretical background subsequently applied to the
analysis of the results that are discussed in Section IV. Finally, our
conclusions are stated in Section V.

\section{The Ising model with LR interactions and the simulation method}

In this paper we present and discuss simulations of the Ising model in $d =
1 $ dimensions, whose Hamiltonian, $H$, is given by

\begin{equation}
H = -J \sum_{\langle i,j \rangle} S_{i} S_{j}/r_{ij}^{d + \sigma},
\label{Hamil}
\end{equation}

\noindent where $J > 0$ is the (ferromagnetic) coupling constant, $S_{i}$
is the spin variable at the site of coordinates i, which can 
assume two values, $S_{i} = \pm 1 $,
the summation is extended to all pairs of spins placed at distances $r_{i,j}
= |r_i - r_j|$, and $\sigma$ is a parameter that controls the decay of 
LR interactions.

Simulations are performed by using samples of length $L\leq 1 \times 10^5$ and taking
periodic boundary conditions. The LR interactions described by the
Hamiltonian of equation (\ref{Hamil}) are evaluated up to a distance 
$|r_i-r_j|=L/2$. Also, simulations with LR interactions truncated at the $Nth$
neighbor, i.e., $J=0$ for $r>N$, have been performed in order to apply a
Finite Range Scaling (FRS) analysis \cite{Glumac}, and the results will be
briefly discussed. Spin update is performed by using the standard Metropolis
dynamics. Also, during a Monte Carlo time step (MCS) all the spins of the
sample are updated once, on average.

In order to carry out the calculations we chose $\sigma =0.75$, because for
this value of the parameter the critical exponents of the Ising model are
expected to be sufficiently different from mean-field values 
($\sigma = 0.50$) to allow a meaningful
comparison with RG results \cite{nag,mon,uze}. Furthermore, one also likes to be as far as
possible from $\sigma =1.00$, where strong Kosterlitz-Thouless behaviour is known
to occur \cite{car}.

During the simulations we recorded the time dependence of the following
observables: (i) The order parameter or average magnetization ($M(t,\tau)$)
given by

\begin{equation}
M(t,\tau)= \frac {1}{L} \langle \sum_{i=1}^{L} S_{i}(t,\tau) \rangle,
\label{magne}
\end{equation}

\noindent where $\tau=\frac{T-T_c}{T_c}$ is the reduced temperature and $T_c$
is the critical temperature.

\noindent (ii) The susceptibility ($\chi(t,\tau)$) evaluated as the
fluctuations of the order parameter, namely

\begin{equation}
\chi (t,\tau )=(M^2(t,\tau )-M(t,\tau)^2),  \label{chi}
\end{equation}

\noindent where $M^2(t,\tau )= \frac {1}{L^2} \langle (\sum_{i=1}^{L} S_{i}(t,\tau))^2
 \rangle$.

\noindent (iii) The autocorrelation of the spin variable

\begin{equation}
A(t,\tau )=\frac 1L\langle \sum_{i=1}^LS_i(t,\tau )S_i(0,\tau )\rangle .
\label{autoc}
\end{equation}

\noindent (iv) The time correlation of two spins separated a distance r at the 
critical point

\begin{equation}
C(t,r)=\frac 1L\langle \sum_{i=1}^LS_i(t)S_{i+r}(t)\rangle .
\label{correlations}
\end{equation}

\noindent (v) The autocorrelation of the order parameter at the critical point,
when the initial condition corresponds to uncorrelated states, given by

\begin{equation}
Q(t) = \frac {1}{L^2} \langle \sum_{i=1}^{L} S_{i}(t) \sum_{i=1}^{L}
S_{i}(0)\rangle.  \label{Qt}
\end{equation}

\noindent (vi) The second-order Binder cumulant ($U(t)$), when the
initial condition corresponds to the ground state, namely,

\begin{equation}
U(t,\tau) = \frac{M^2(t,\tau)}{M(t,\tau)^2} -1 ,  \label{binder}
\end{equation}

\noindent where in all cases the brackets indicate configurational averages
performed over a number $n_s$ of different samples started from equivalent 
(but different in the case of $T=\infty$) initial conditions.
 
\section{Brief theoretical background}

\textit{Short-time dynamics (STD)}: Let us now analyse the expected
short-time dynamic behaviour when the system starts from a disordered
(uncorrelated) configuration, but with a small initial magnetization.
According to the argument of Janssen et al. \cite{jan}, the general scaling
approach of the order parameter for the nonconservative dynamics of model A
(according to the classification of Hohenberg and Halperin \cite{hoha}), is
given by

\begin{equation}
M(t,\tau,L,M_{0}) = b^{-\beta/\nu} M^{~}(t/b^{z}, b^{1/\nu}\tau, L/b,
b^{x_{0}}M_{0}),  \label{escamag}
\end{equation}

\noindent where $b$ is a scaling parameter, and $\beta$ and $\nu$ are the order
parameter and correlation length (static) critical exponents, respectively.
Also, $z$ is the dynamic exponent. Furthermore, $x_{0}$ is a new exponent,
introduced by Janssen et al \cite{jan}, which accounts for the scaling
dimension of the initial magnetization $M_{0}$, in the $M_{0} \rightarrow 0$
limit.

For sufficiently large lattices, at the critical point ($\tau\equiv 0$), and by
setting $b=t^{1/z}$, equation (\ref{escamag}) becomes

\begin{equation}
M (t, M_{o}) = t^{-\beta/ \nu z} M(t^{\frac{x_{o}}{z}}M_{o}),  \label{mcero}
\end{equation}

\noindent which holds for a time short enough such that the correlation
length ($\xi(t) \propto t^{1/z}$) is not so large ($\xi \ll L$).
Furthermore, for times even shorter than the crossover time ($t_{x} \approx
M_{o}^{-z/x_{o}}$), but larger than the microscopic time that is set when
the correlation length is of the order of a single lattice spacing, equation 
(\ref{mcero}) becomes

\begin{equation}
M(t) \propto M_{0}t^{\theta},  \label{in_in}
\end{equation}

\noindent which describes the (power-law) initial increase of the
magnetization with exponent $\theta = x_{0}/z - \beta/ \nu z$.

In the absence of an initial magnetization ($M_{0}\equiv 0$), and at
criticality, the scaling behaviour of the susceptibility is given by

\begin{equation}
\chi(t) \propto t^{\gamma/ \nu z},  \label{scachi}
\end{equation}

\noindent where $\gamma$ is the susceptibility exponent. Also, under these
conditions ($\tau = 0$ and $M_{0} = 0)$, the time autocorrelation function
is expected to follow a power law with time according to

\begin{equation}
A(t)\propto t^{-\lambda},  \label{auto}
\end{equation}

\noindent where the critical exponent is given by $\lambda = d/z-\theta$,
i.e., even in the absence of an initial magnetization, $\lambda$ depends on
the exponent $\theta$ that describes the initial increase of the order
parameter according to equation (\ref{in_in}).

On the other hand, by starting with randomly generated configurations,
the correlation function of the total magnetization is also expected to
follow a power law with time according to

\begin{equation}
Q(t)\propto t^{\theta},  \label{Q}
\end{equation}
\noindent i.e., a relationship that allows us to obtain the initial increase
exponent avoiding the numerical extrapolation $M_0\rightarrow0$ \cite{Oliveira}.

Finally, the two-spins time correlation allows us to obtain an independent 
determination of dynamic exponent $z$ by mean of the following scaling form\cite{Bray}
\begin{equation}
C(t,r)=r^{-(d-2+\eta)} C�(r/\xi(t)),  \label{C}
\end{equation}

\textit{Standard relaxation dynamics (SRD)}. STD measurements can be further
reinforced by independent measurements of the SRD, which are started from a
fully ordered or ground state configuration and are performed at
criticality. In this way, one could be able not only to test the validity of
some exponents evaluated by means of the STD method, as well as the critical
temperature, but also obtain additional exponents
and test the validity of relationships between them, e.g., the hyperscaling
relationship \cite{Zheng}. In fact, by starting from a ground-state
configuration with all spins pointing in the same direction ($T=0$), upon
annealing to criticality, the SRD scaling approach is given by (see also 
equation (\ref{mcero})) 

\begin{equation}
M(t,\tau,L) = b^{-\beta/ \nu} M(t/b^z, b^{1/ \nu} \tau, L/b).  \label{que}
\end{equation}
For large lattices and by setting $b=t^{1/z}$, this dynamic scaling form
leads to

\begin{equation}
M(t,\tau )\propto t^{-\beta /\nu z}M(t^{1/\nu z}\tau ).  \label{quen}
\end{equation}
It is well known that this power-law decay of the order parameter is valid
within the long-time regime, but several numerical results indicate that it
also holds in the short-time regime.

On the other hand, by taking the logarithmic derivative of equation (\ref
{quen}) with respect to the reduced temperature, evaluated at the critical
point, one gets

\begin{equation}
\frac{\partial log M(t,\tau)}{\partial \tau}\vert_{\tau=0} \propto t^{1/ \nu
z},  \label{deri}
\end{equation}

\noindent which allows us to evaluate the exponent $1/\nu z$, by
performing measurements at and slightly away from the critical point.
Furthermore, just at the critical point the second-order Binder cumulant is
expected to behave according to

\begin{equation}
U(t)\propto t^{d/z}.  \label{cumu}
\end{equation}
\thinspace

It is worth mentioning that because of the small 
nonequilibrium correlation length for short-ranged models
both STD and SRD are free of finite size effects. 
However, in long-ranged models finite-size
effects also appear due to the fact that the finite size
yields to a truncated interaction range. So,
this effect remains even during the short-time regime investigated in this paper
and it is worth knowing its influence on both the critical
temperature and the critical exponents. 

\section{Results and discussion.}

\subsection{Standard relaxation dynamics}

\begin{figure}[h]
\begin{center}
\includegraphics[width=0.9\textwidth,angle=0]{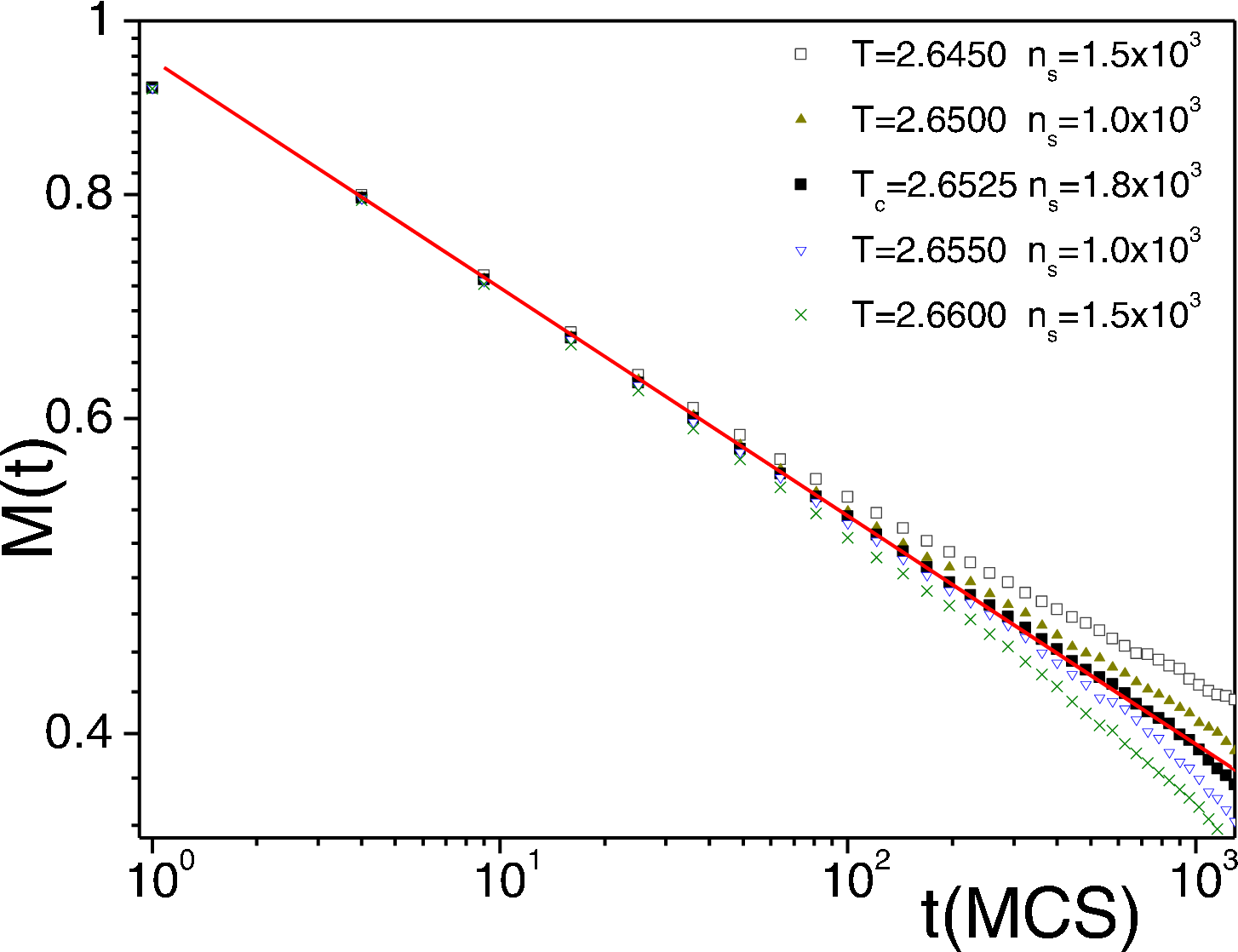}
\end{center}
\caption{(colour online) Log-log plots of the time evolution 
of the magnetization $M(t)$ obtained after annealing from $T=0$ 
(ground state) to the indicated temperatures. Data corresponding 
to the system size $L=2\times10^4$. The solid line shows the fit of the curve 
obtained for $T_c=2.6525$, according to equation (\ref{quen}). 
The number of averaged configurations ($n_s$) is also indicated. 
More details in the text. }
\label{Mvsmcs}
\end{figure}

Focusing our attention first on the relaxation dynamic behaviour at
criticality, figure \ref{Mvsmcs} shows the time evolution of the
magnetization at different temperatures for the system size $L=2\times10^4$.
For this system size the \textit{critical} temperature $T_c=2.6525(25)$ 
was found by searching the smallest standard deviation from the power 
law (equation (\ref{quen})), and the error bars were assessed by considering 
closest temperatures that present noticeable but small deviations. Also, 
from the fit of the data the critical exponent $\beta /\nu z=0.129(6)$ 
was determined. 

\subsubsection{Finite-size effects}

\begin{figure}[h]
\begin{center}
\includegraphics[width=0.8\textwidth,angle=0]{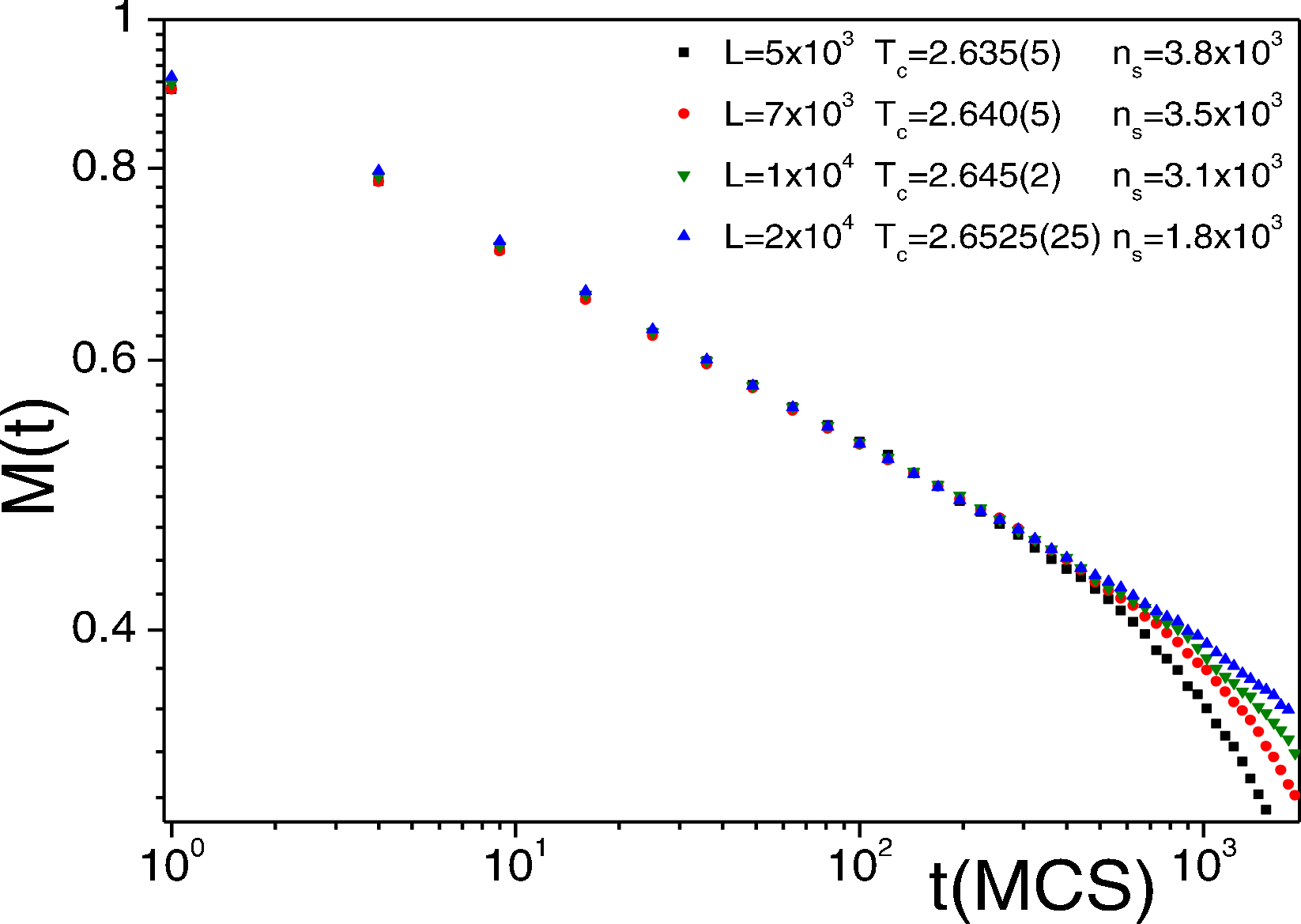}
\end{center}
\caption{(colour online) Log-log plot of the time evolution of the magnetization $M(t)$
obtained after annealing from $T=0$ (ground state) to the critical temperatures
corresponding to the indicated system sizes ($L$). The number of averaged 
configurations ($n_s$) is also indicated. More details in the text. }
\label{M1}
\end{figure}

\begin{figure}[h]
\begin{center}
\includegraphics[width=0.8\textwidth,angle=0]{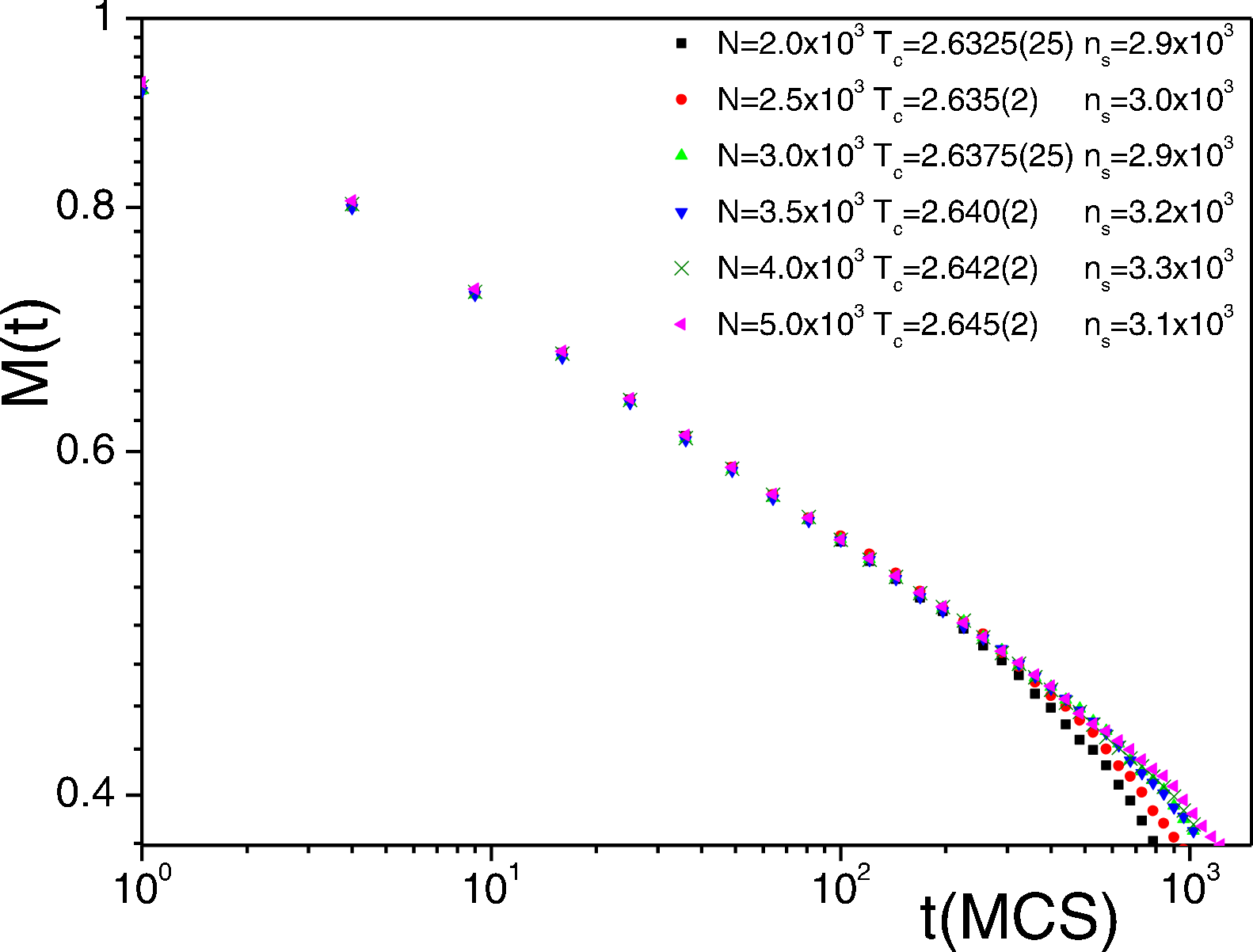}
\end{center}
\caption{(colour online) Log-log plot of the critical relaxation of the magnetization $M(t)$
from $T=0$ for the system size $L=10^4$ and different interaction ranges $N$. 
The number of averaged configurations ($n_s$) and effective critical temperatures are
also indicated. More details in the text. }
\label{L10000Ns}
\end{figure}

\begin{figure}[h]
\begin{center}
\includegraphics[width=0.8\textwidth,angle=0]{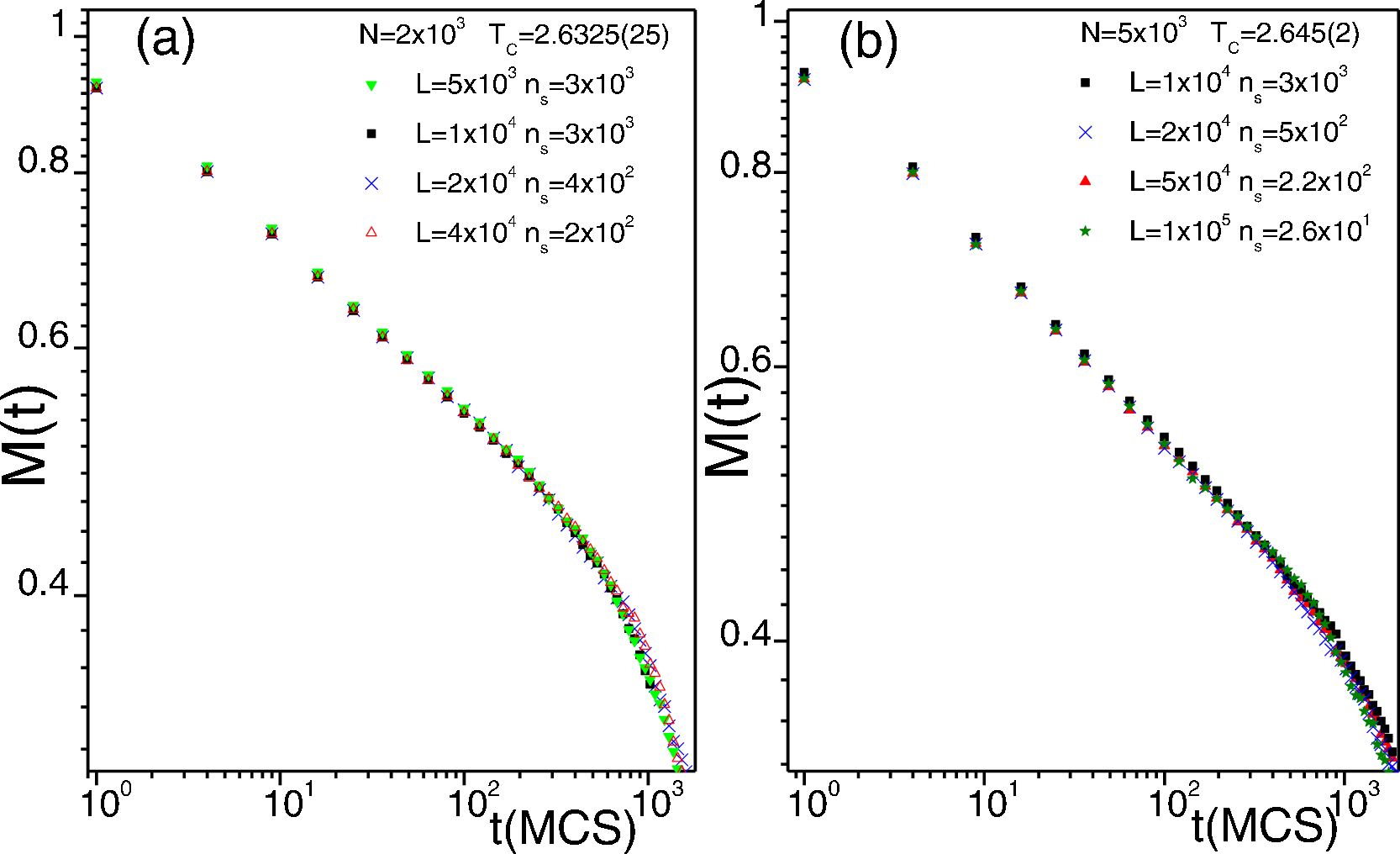}
\end{center}
\caption{(colour online) Log-log plot of the critical relaxation of the magnetization $M(t)$
from $T=0$ for the indicated system sizes $L$ and fixed interaction 
range a) $N=2\times10^3$ and b) $N=5\times10^3$. The number of averaged configurations ($n_s$) 
and effective critical temperatures are also indicated. More details in the text. }
\label{NLs}
\end{figure}

In order to investigate the influence of finite-size effects on the results, 
the procedure described above was carried out not only for several system 
sizes (see figure \ref{M1}), but also for different interactions ranges. 
The purpose of that type of study is to distinguish between two different sources of size effects:
those caused by the finiteness of the sample and those other caused by the finite-interaction range. In fact, in contrast to
the case of results often obtained by using models with short-range interactions \cite{RPP}, here the expected 
power-law behaviour of the physical observables is observed for temperatures that depend on the size, i.e. effective
critical temperature. Then it is possible to understand this situation as an additional size effect that is caused 
by the truncated-interaction range of the long-range interaction rather than by the usual finite number of spins sites considered
in Monte Carlo simulations. Indeed, a finite system-size sample implies a truncated-interactions range, i.e., the maxima number of 
neighbours ($N_{max }$) at each side of the central spin considered in order to evaluate the Hamiltonian given by equation
(\ref{Hamil}) is finite, and due to the periodic boundary conditions used, one has $N_{max}=L/2$. Following that, simulations
with different $N \le N_{max}$ and $L$ values were carried out. Figure \ref{L10000Ns} shows the critical relaxation of the magnetization for a system size $L=10^4$ and
different $N$ values. The data indicate that the effective critical temperature and range of the power-law behaviour depends on the value of $N$ but the corresponding critical exponent remains unaffected,
within the short-time regime.  Furthermore, this statement is reinforced by the results shown in figures \ref{NLs} (a) and (b) that correspond to $N=2\times10^3$ and $N=5\times10^3$ and 
different $L$ values, respectively. 
Summing up, the (almost) perfect overlap of the curves observed within the suitable time interval defined by
each system size shows that in the ILR model the critical temperature must be changed with the sizes
meanwhile the critical exponents are no longer influenced.

\subsubsection{Finite-Range Scaling (FRS) Analysis}

A FRS analysis has also been applied in order to obtain the critical temperature 
in infinite interaction range (thermodynamic limit). 
This type of analysis has already been developed by analogy with the finite-size scaling method
\cite{Glumac}. The basic idea behind this approach is to study systems with different truncated-interaction
ranges and obtain information on the critical behaviour by mean of scaling properties. In this way, 
based on \cite{Glumac}, the following scaling dependence has been proposed, 

\begin{equation}
T_c(N)=T_c(\infty )+A/N^{x_T}, \label{TcN}
\end{equation}

\begin{figure}[h]
\begin{center}
\includegraphics[width=0.8\textwidth,angle=0]{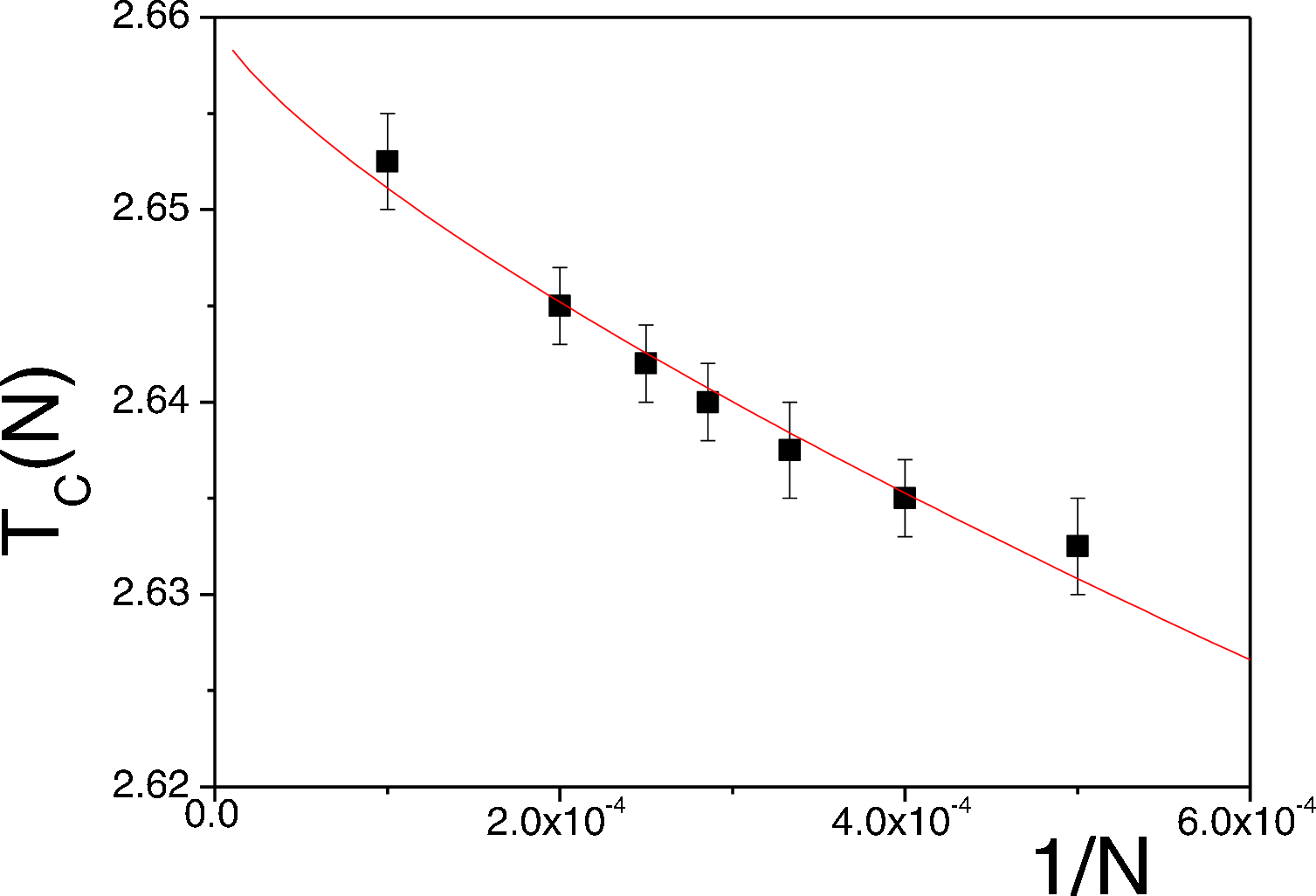}
\end{center}
\caption{(colour online) Plot of the effective critical 
temperature at function of the inverse of the interaction range (N).
The continuous line corresponds to the fit performed with the aids of equation (\ref{TcN}). More details in the text.}
\label{Tc(N)}
\end{figure}

\noindent where $T_c(\infty )$ is the critical temperature for the infinite
interaction range, $x_T$ is the convergence exponent, and $A$ is a constant.
Figure \ref{Tc(N)} shows the obtained $T_c(N)$ values at function of $N^{-1}$, which was
fitted with the aid of equation (\ref{TcN}) (continuous line), getting a value for the fitted convergence
exponent ($x_T=0.74(2)$), and the critical temperature ($T_c(\infty )=2.660(4)$). The obtained critical
temperature by this approach interpolates between the previously reported values for $\sigma =0.70$ ($T_c(\infty )=2.929$ \cite{Glumac} and 
$T_c(\infty )=2.9269$ \cite{Barati}), and for $\sigma =0.80$ ($T_c(\infty )=2.431$ \cite{Glumac} and $T_c(\infty )=2.4299$ \cite{Barati}), which were obtained
by means of analytic calculations with the transfer matrix method and FRS analysis.

\subsubsection{Critical Exponents}

\begin{figure}[h]
\begin{center}
\includegraphics[width=0.7\textwidth,angle=0]{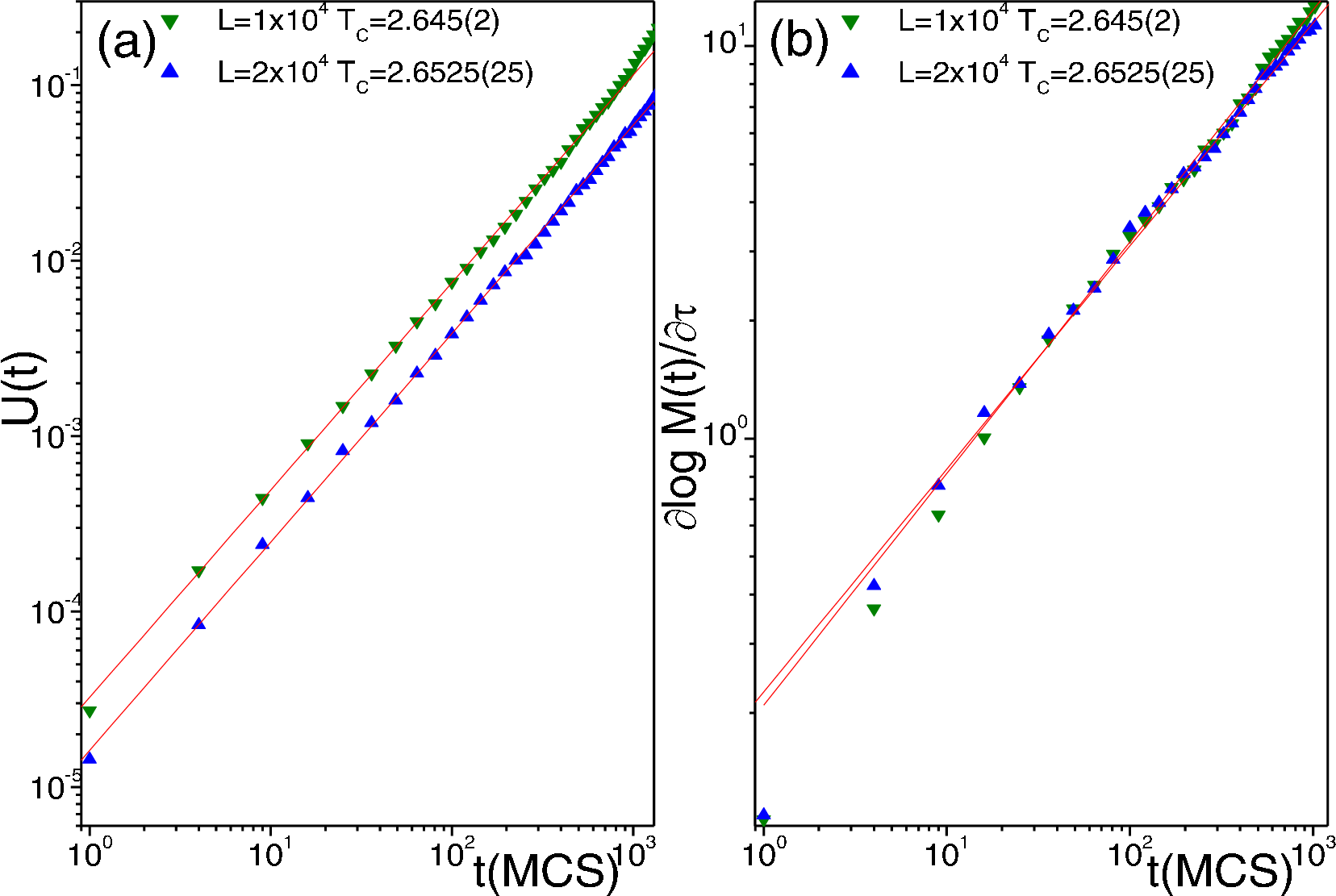}
\end{center}
\caption{(colour online) Time evolutions of dynamic observables obtained after annealing at 
effective critical temperature from $T=0$ (a) The second-order Binder cumulant ($U(t)$), and (b) the
logarithmic derivative of the magnetization with respect to the reduced
temperature ($\frac{\partial Log(t)}{\partial \tau}$). The solid lines
indicate the fits performed with the aid of equations (\ref{deri}) and (\ref{cumu}), respectively. 
The system sizes ($L$), and the corresponding effective critical temperatures ($T_c$),
are also indicated. }
\label{Uder}
\end{figure}

The already discussed results suggest that the system size $L=10^4$ 
is large enough for the evaluation of the critical exponents 
within a suitable time interval, namely $(10,900)$MCS. In order to
verify the above statement and to obtain the complete set of critical 
exponents, the SRD of the physical observables was obtained for 
system sizes of $L=10^4$ and $L=2\times10^4$ until $10^3$ MCS.
Figure \ref{Uder} (a) shows the time evolution of
the second-order Binder cumulant at the effective critical temperature
that can be fitted with a power law with the exponents 
listed in Table \ref{table1} (3th column). From these values the dynamic 
exponent $z$ was estimated to be close to $z=0.84(2)$ (see Table \ref{table1} 
5th column), e.i. a figure that is significantly larger than the RG results, given by 
$z_{RG}=0.775$ \cite{chen}. In principle one could 
expect that this disagreement may be most likely  
due to the fact that $z$ depends on the specific dynamics used,
as in the case of the short-ranged Ising model \cite{Okano}. 
Nevertheless, this discrepancy could also be attributed to an 
underestimation of the RG calculation, again as in the case of 
the short-ranged Ising model\cite{Halperin}. 
On the other hand, by using measurements of the magnetization performed 
at two adjacent temperature points of the effective critical
one, the logarithmic derivative of the magnetization with respect to the reduced temperature was 
obtained. Figure \ref{Uder}(b) shows that this observable also exhibits 
a power-law behaviour and the fitted exponents are listed in Table 
\ref{table1}, 4th column. Furthermore, by replacing the obtained value of $z$ 
in the exponent corresponding to the logarithmic derivative, one 
gets $1/\nu =0.48(2)$ (see Table \ref{table1} 6th column), 
in agreement with both the RG prediction, namely, $1/\nu =0.4765$
\cite{chen,Binder}, and with Monte Carlo simulations performed at
equilibrium, $1/\nu =0.469$ \cite{Binder}.
\vspace{1cm}

\begin{table}[htbp]
\begin{tabular}{|c|c|c|c|c|c|c|}
\hline
 $L$            & $\beta/\nu z$ & $d/z$ & $1/\nu z$ & $z$     &$1/\nu$   & $\beta/\nu$ \\
\hline
$ 1\times10^4 $ & 0.129(7)      &1.20(2)& 0.59(2)   & 0.83(1) &0.49(2)  & 0.107(5)    \\
$ 2\times10^4 $ & 0.129(6)      &1.19(3)& 0.57(2)   & 0.84(2) &0.48(2)  & 0.109(6)    \\
RG              &               &       &           & z=0.775 &0.4765   & 0.125         \\
\hline
\end{tabular}
\caption{List of exponents obtained by means of SRD measurements of the magnetization ($\beta/\nu z$), 
Binder Cumulant ($d/z$), and logarithmic derivative of the magnetization with respect to the reduced temperature 
($1/\nu z$). The estimated critical exponents $z$, $1/\nu$, and $\beta/\nu$, as well as the RG predictions
are also listed for the sake of comparison.}
\label{table1}
\end{table}

It is worth to mentioning that the error bars of the evaluated exponents 
are not easy to estimate because they are introduced by several sources such as 
insufficient statistics, arbitrariness in the time interval used to fit 
the power-law behaviour of the observables, and finally the use of an approximate 
effective critical temperature, $T_c$. 
In order to have an estimation of the magnitude of the error due to the former
source, a variant of the blocking method was used \cite{Newman}. For this purpose
one proceeds as follows: the time dependence of each observable is fitted
for several independent sets of measurements, then, the error bars are obtained 
by accounting for the spreading of the obtained values.
In the case of the time interval used for the power-law fit, 
we found that the selection of the microscopic time accounts for the major
error. So, the reported exponents correspond to a fixed microscopic time that is
established after the first 10MCS, and the error bars include the values obtained 
by taking microscopic times within the range $10 - 100$MCS. On the other hand, 
the error due to the approximate critical temperature cannot be estimated directly.

\subsection{Short-Time Dynamics}

\begin{figure}[ht]
\begin{center}
\includegraphics[width=0.8\textwidth,angle=0]{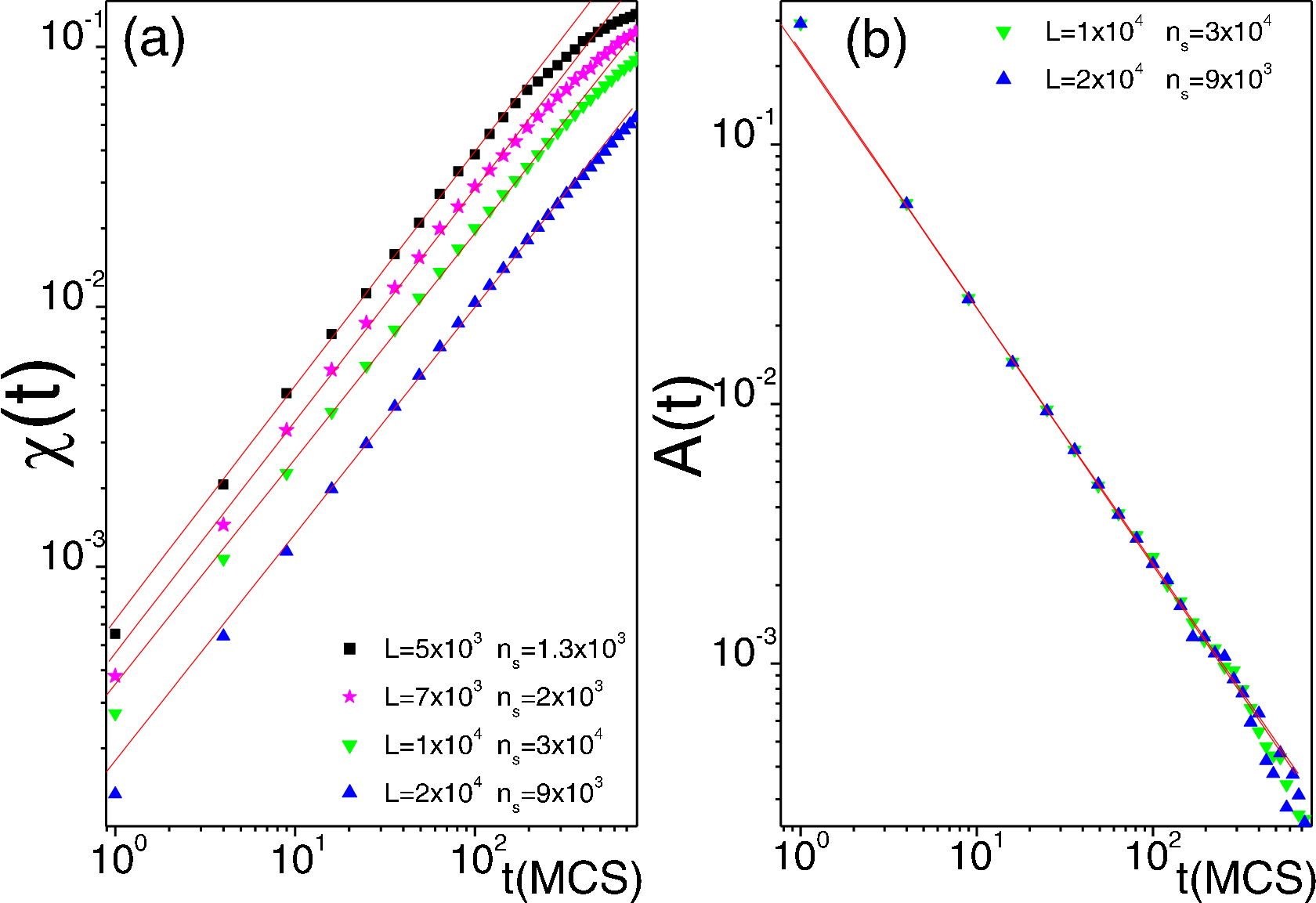}
\end{center}
\caption{(colour online) Time evolution measured after quenching from uncorrelated (disordered)
states to the corresponding effective critical temperature $T_C$ of 
(a) the susceptibility $\chi(t)$ and (b) the autocorrelation $A(t)$. 
The solid lines indicate the fits performed with the aid of equations
(\ref{scachi}) and (\ref{auto}), respectively. 
The number of averaged configurations ($n_s$) and system sizes ($L$) are also indicated.}
\label{M_A}
\end{figure}

Now we turn our attention to the STD measurements. 
The STD evolution exhibits a weak dependence on the quenching 
temperature, so this shortcoming hinders an independent estimation 
of $T_c$. Consequently, in the simulations we used the values obtained
from SRD measurements. As in that case, a finite-size analysis of the time evolution of the susceptibility 
(see figure \ref{M_A}(a)) allows us to determine the suitable time 
interval used in order to perform the fitting procedure. 
In this way, for the system size $L=10^4$ 
the power-law behaviour is observed until 400MCS. Also, 
the autocorrelation function (figure \ref{M_A}(b)) exhibits a power-law decay 
at the same time interval. The critical exponents $\gamma /\nu z$ and $\lambda$ 
obtained by means of a fit with the aids of equations (\ref{scachi}) and (\ref{auto}), respectively, 
are presented in Table \ref{table2}. The error bars of the critical 
exponents were estimated in the same way as for the case of the SRD measurements, 
and they include the values corresponding to microscopic times taken 
from the interval $4 -36$MCS. 

\begin{table}[htbp]
\begin{tabular}{|c|c|c|c|c|c|c|}
\hline
\textbf{L}    &$\gamma/\nu z$ & $d/z-\theta$ & $\theta $ & $z$ &$\gamma/\nu$& $\beta/\nu$ \\
\hline
$ 1\times10^4 $ &0.87(2)     &0.99(1)       & 0.200(5)  &0.840(8)&0.73(2)  &0.13(1)      \\
$ 2\times10^4 $ &0.88(1)     &0.99(1)       & 0.201(4)  &0.839(8)&0.74(1)  &0.130(9)      \\
RG              &             &              & 0.2171    &0.775   &0.75    &0.125         \\
\hline
\end{tabular}
\caption{Critical exponents obtained from the STD evolution 
of the susceptibility ($\gamma/\nu z$), autocorrelation ($d/z-\theta$) and initial increase
of the magnetization ($\theta$). The calculated exponents $z$, $\gamma/\nu$ and $\beta/\nu$ 
and the corresponding RG predictions are also included.}
\label{table2}
\end{table}

\begin{figure}[ht]
\begin{center}
\includegraphics[width=0.8\textwidth,angle=0]{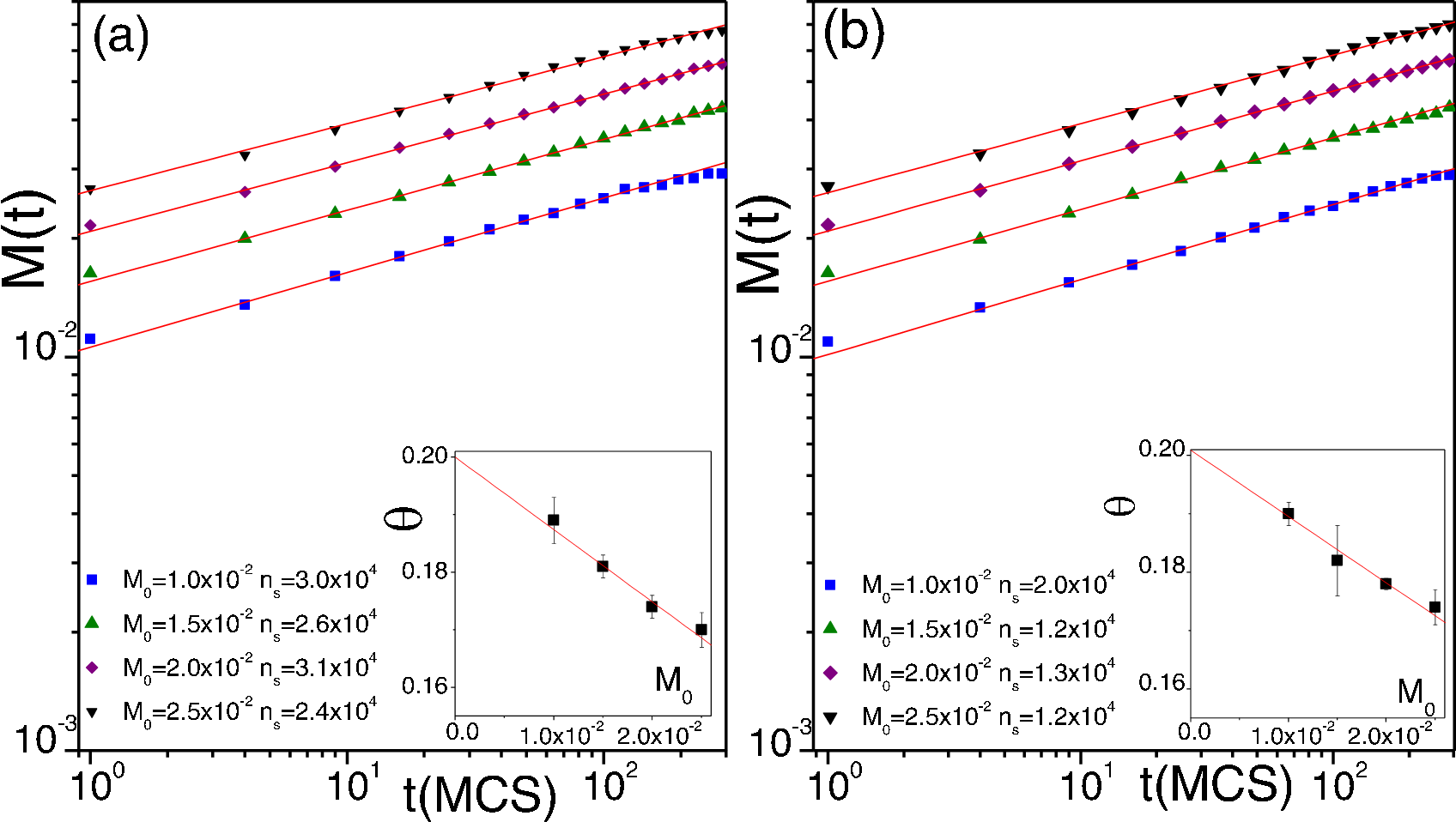}
\end{center}
\caption{(colour online) Log-log plot of $M(t)$ versus time showing 
the initial increase of the magnetization obtained after quenching the system from uncorrelated 
(disordered) states, with a small magnetization $M_0$, to $T_c$. 
The data correspond to systems sizes (a) $L=10^4$ and (b) $L=2\times10^4$. 
The solid lines show the fits obtained according to equation 
(\ref{in_in}). The inset shows the linear extrapolation of the values of the exponent 
to $M_0\rightarrow0$. The number of averaged configurations ($n_s$)
is also indicated.}
\label{theta}
\end{figure}

In contrast with these measurements performed by
setting $M_0\equiv 0$, the initial increase of the magnetization has to be
measured for vanishingly small values of $M_0$, as is shown in the 
figures \ref{theta} (a) and (b) for system sizes $L=10^4$ and $2\times10^4$, respectively. 
Note that the simulation time verifies that $t\ll t_x$.
The insets show the power-law exponents obtained by the fit by means of 
equation (\ref{in_in}) and the extrapolation for $M_0\rightarrow 0$.
This procedure yields $\theta$ values reported in Table \ref{table2} (4th column)
which are close to the RG prediction\cite{chen}. 
Now, by using the relationship $\lambda =d/z-\theta $ and replacing the determined 
exponents, one gets the dynamic exponent $z$ (see Table \ref{table2}, 5th column). 
The obtained value $z=0.84$ is consistent with our previous SRD determinations but slightly 
higher to the RG result ($z_{RG}=0.775$)\cite{chen}. Also, it interpolates 
between previously published STD results corresponding to a system 
of size $L=3000$, which are given by $z=0.81(1)$ and $0.96(4)$, 
for $\sigma =0.70$ and $0.80$, respectively\cite{loca}. 
On the other hand, one can use the values of both $\gamma /\nu z$ and $z$ in order
to estimate $\gamma /\nu)$ (see Table \ref{table2}, 6th column). Furthermore, by assuming that the
hyperscaling relationship ($d-2\beta/\nu = \gamma /\nu$) holds, one
can obtain the STD estimation of $\beta /\nu =0.130(9)$.
It is worth to mention that RG calculations obtained from the asymptotic expansion in 
$\epsilon=2\sigma-d$ up to second order yield $\eta=2-\sigma$=1.25\cite{fish}. 
Then, by using the standard scaling relationships $\gamma/\nu=2-\eta$ and 
$\beta/\nu=(d-2+\eta)/2$, the exponents $\gamma /\nu =\sigma =0.75$ 
and $\beta /\nu =\frac{d-\sigma }2=0.125$ can be obtained in excellent 
agreement with our STD estimations. 

\begin{figure}[th]
\begin{center}
\includegraphics[width=0.9\textwidth,angle=0]{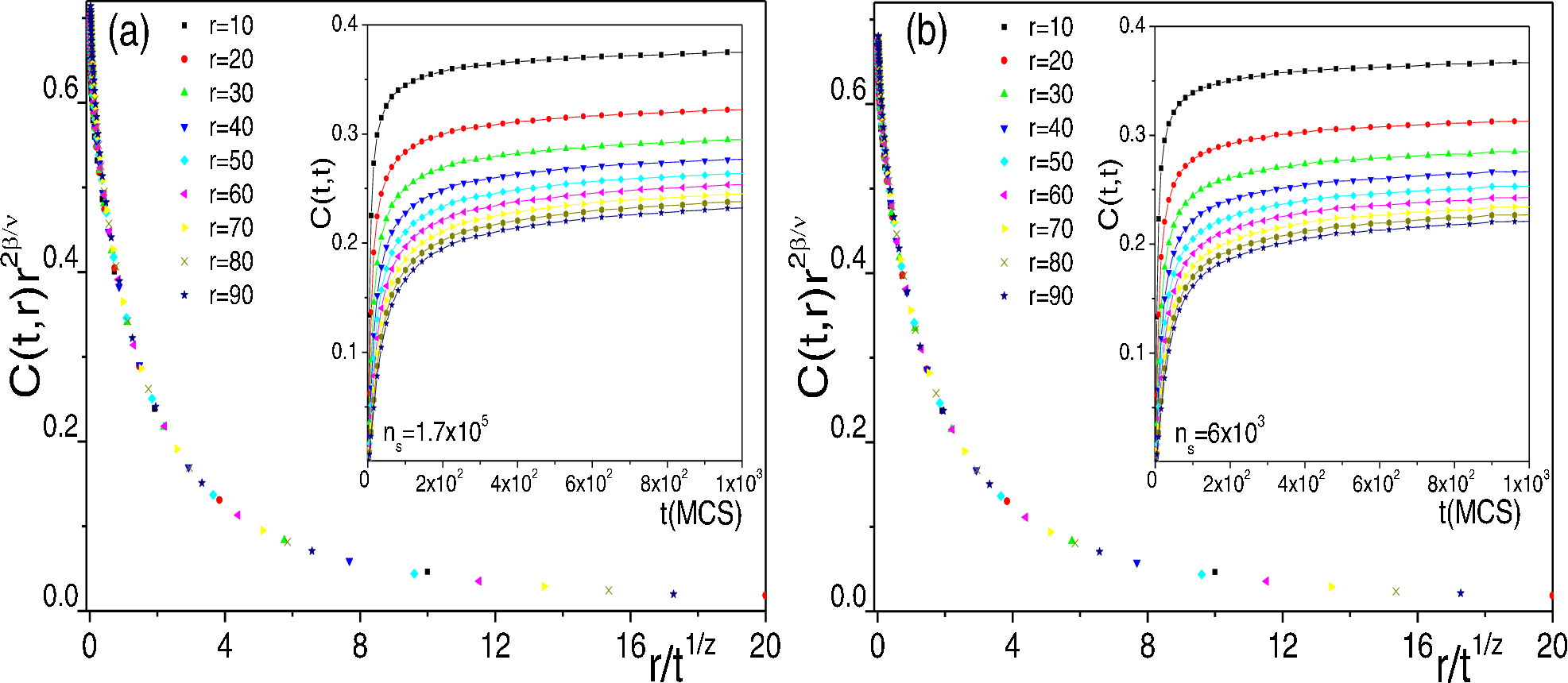}
\end{center}
\caption{(colour online) Plots of the scaled spin-spin correlation function 
$ r^{2\beta\nu} C(r,t)$ as a function of the scaled 
variable $x = r/t^{1/z}$, as obtained for (a) $L=10^4$ and (b) $L=2\times10^4$.
The insets show the time evolution of $C(t,r)$ for the indicated $r$ values 
after quenching randomly generated configurations to $T_c$. The collapses 
show in the main panels were obtained using by $z=0.84$ and $\beta/\nu=0.125$. 
The number of averaged configurations ($n_s$) are also indicated.}
\label{ctt}
\end{figure}

Furthermore, just by starting with random configurations and
measuring the autocorrelation function of the magnetization ($Q(t)$)
given by equation (\ref{Q}), one can also obtain the initial
increase exponent $\theta =0.180(6)$, as shown in figure (\ref{Q}). 
Due to the fact that in this case the fluctuations are more 
pronounced, the calculation of the correlation function requires 
better statistics and consequently the simulations were done up to $200$MCS
for $L=10^4$. The error bars include the figures obtained for 
microscopic times within the range $4 -36$MCS. The value of the 
exponent $\theta$ is close to previous measurement obtained by using 
the numerical extrapolation $M_0\rightarrow 0$, 
namely, $\theta =0.201(4)$. Furthermore, by using
this independent estimation of $\theta$ and applying the
previously described procedure, the exponents
$z=0.855(9)$, $\gamma /\nu =0.74(2)$, and $\beta /\nu
=0.13(1)$ can be obtained, which of course, are in good agreement
with our previous estimations.
\begin{figure}[th]
\begin{center}
\includegraphics[width=0.5\textwidth,angle=0]{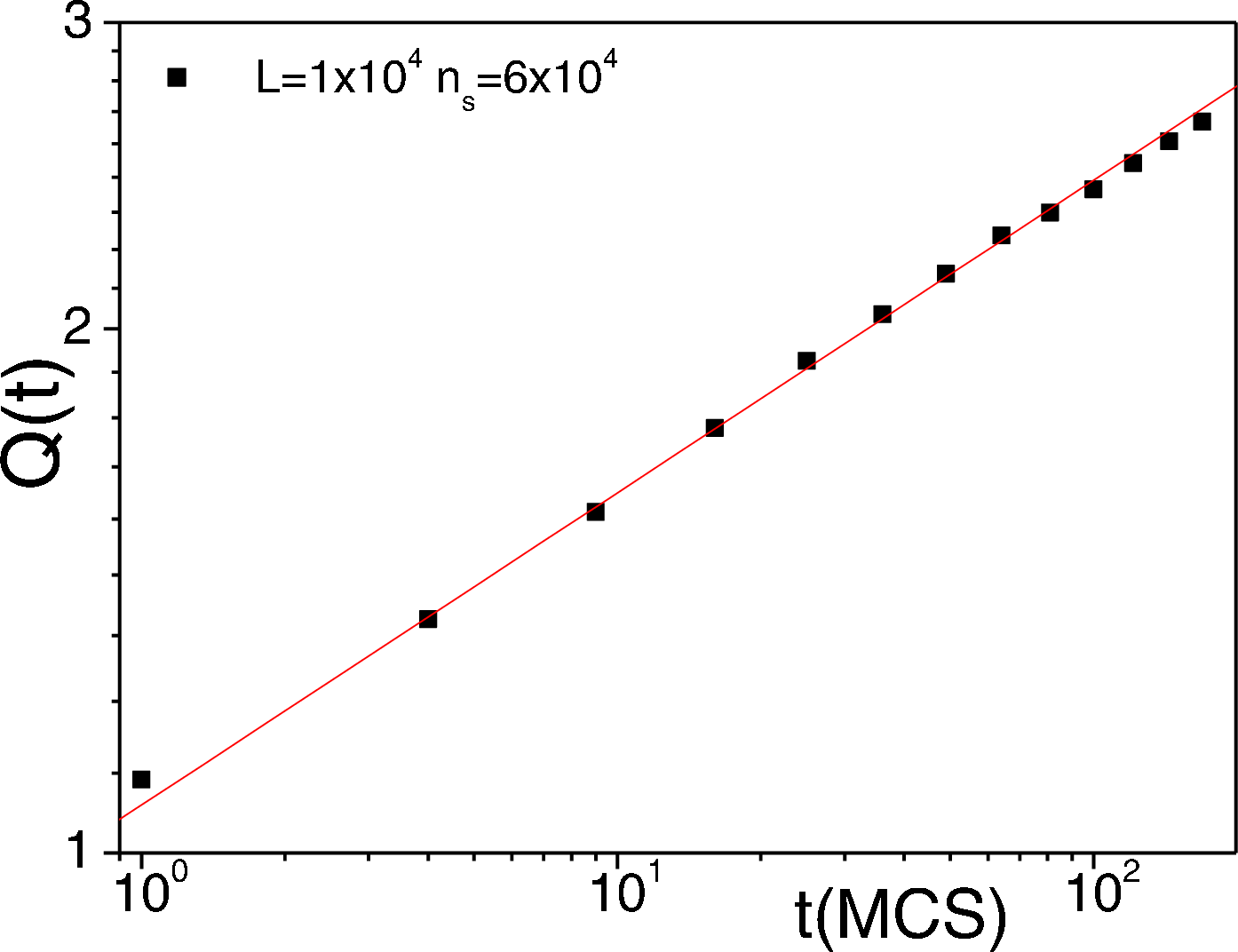}
\end{center}
\caption{(colour online) Log-log plot of the time evolution of the autocorrelation function of
the magnetization after quenching randomly generated configurations to 
$T_c=2.645$. The solid line shows the fit performed with the aid of equation (\ref{Q}). The number
of averaged configurations ($n_s$) and system size ($L$) are also indicated.}
\label{Q}
\end{figure}
On the other hand, in order to obtain an additional independent estimation of the dynamic exponent $z$,
the scaling behaviour of the spin-spin correlation functions ($C(t,r)$) 
was studied for different values of $r$ ranging from 10 to 90 (see insets
of figure \ref{ctt}). The main panels of figure \ref{ctt} show the 
best collapse of the $C(r,t)$ obtained by using the conventional critical
scaling (equation (\ref{correlations})) and assuming that the hyperscaling relation 
($d= 2\beta /\nu + \gamma/\nu $) and $\eta=2-\gamma/\nu $ hold. From these results,
the exponents $z=0.84(2)$ and $\beta/\nu=0.125(3)$ were obtained .
The error bars were determined by considering the values where noticeable 
deviations from the collapsed form were observed (not shown here for the sake of space). 
These results are in excellent agreement with our previous determinations and further support
the self-consistence of the obtained results by means of different dynamical methods.

\section{Conclusions}

In this paper we present and discuss the results of extensive simulations of
the non-equilibrium dynamic behaviour of the LR Ising magnet with interactions decaying as 
$r^{-(d+\sigma )}$, in $d=1$ dimensions and with $\sigma =0.75$. 

Power-law behaviour of the relevant observables was found at temperatures 
which depend on the interaction range, for both the relaxation and the short-time regimes. 
The results allow us to verify that the finite-sizes effects only affects both the effective 
critical temperature and the time power-law range, while in contrast the critical exponents 
remaining inalterable, within the studied of interaction ranges. 

Furthermore, finite range scaling analysis was applied in order to obtain the critical 
temperature in the thermodynamic limit which yields $T_c(\infty)=2.660(4)$. It is found that 
all the estimated static critical exponents ($\gamma /\nu $, $\beta /\nu $, and $1/\nu)$ are
in good agreement with RG results. Also, the dynamic exponent of the STD initial increase of the 
magnetization ($\theta$) is close to RG results. The estimations of the dynamic exponent ($z$) of 
the time correlation length from SRD and STD measurements are in agreement, but they are slightly 
different from the RG results. This difference would be due to insufficiency of the two-loop 
expansion in RG analysis, or it may also be a consequence of a dependence on the specific 
Monte Carlo dynamics used (Metropolis in the present paper).

Summing up, the reported results lead us to conclude that the comparison between 
both types of dynamic measurements, annealing and quenching, provides relevant information 
on the critical behaviour of a system with long-range interactions, allowing 
the evaluation of both dynamic and static critical exponents.


This work was supported financially by CONICET, UNLP, and ANPCyT (Argentina).

\end{document}